\documentclass[preprint,
superscriptaddress,
 amsmath,amssymb,
 aps, citeautoscript,
prb,
]{revtex4-1}

\usepackage{graphicx}
\usepackage{dcolumn}
\usepackage{bm}

\begin{document}
\preprint{APS/123-QED}

\title{Predicting bandgaps and band-edge positions of oxide perovskites using DFT and machine learning}

\author{Wei Li}
\affiliation{Department of Materials Science \& Engineering, University of Delaware, Newark, DE 19716, USA}
\affiliation{Computer, Computational, and Statistical Sciences Division, Los Alamos National Laboratory, Los Alamos, NM, 87545, USA}

\author{Zigeng Wang}
\affiliation{Computer Science and Engineering Department, University of Connecticut, CT 06269, USA}

\author{Xia Xiao}
\affiliation{Computer Science and Engineering Department, University of Connecticut, CT 06269, USA}

\author{Zhiqiang Zhang}
\affiliation{Department of Physics, University of Delaware, Newark, DE 19716, USA}

\author{Anderson Janotti}
\affiliation{Department of Materials Science \& Engineering, University of Delaware, Newark, DE 19716, USA}
\email{janotti@udel.edu}

\author{Rajasekaran Sanguthevar}
\affiliation{Computer Science and Engineering Department, University of Connecticut, CT 06269, USA}

\author{Bharat Medasani}
\affiliation{Delaware Energy Institute, University of Delaware, Newark, DE 19702, USA}
\affiliation{Princeton Plasma Physics Laboratory, Princeton, NJ 08540}
\email{bmedasan@pppl.gov}

\begin{abstract}

Density functional theory within the local or semilocal density approximations (DFT-LDA/GGA) has become a workhorse in the electronic structure theory of solids, being extremely fast and reliable for energetics and structural properties, yet remaining highly inaccurate for predicting bandgaps of semiconductors and insulators. Accurate prediction of bandgaps using first-principles methods is time-consuming, requiring hybrid functionals, quasi-particle GW, or quantum Monte Carlo methods.  Efficiently correcting DFT-LDA/GGA bandgaps and unveiling the main chemical and structural factors involved in this correction is desirable for discovering novel materials in high-throughput calculations.  In this direction, we use DFT and machine learning techniques to correct bandgaps and band-edge positions of a representative subset of ABO$_3$ perovskite oxides. Relying on the results of HSE06 hybrid functional calculations as target values of bandgaps, we find a systematic bandgap correction of $\sim$1.5 eV for this class of materials, where $\sim$1 eV comes from downward shifting the valence band and $\sim$0.5 eV from uplifting the conduction band. The main chemical and structural factors determining the bandgap correction are determined through a feature selection procedure. 

\end{abstract}

\maketitle

\section*{Introduction}
 
The bandgap and band-edge positions (i.e., ionization energy and electron affinity) are basic properties of semiconductors and insulators, and often dictate the suitability of materials for device applications. Their prediction, based on first-principles methods, is key to novel materials discovery. DFT calculations\cite{Kohn1964,Kohn1965} based on LDA\cite{Hohenberg1964} or GGA\cite{Perdew1996, Perdew_metaGGA} are often used to predict stable crystal structures, with lattice parameters within 1-2\% of the experimental values\cite{Tran2016,Lianhua2014}. These calculations are extremely fast and scalable, permitting the study of the energetic and structural properties of thousands of materials with relatively modest computing resources and in relatively short times, playing a central role in current materials discovery research efforts based on high-throughput computation. However, when standard LDA or GGA functionals are employed, bandgaps ($E_g$) predicted by DFT are severely underestimated in comparison to experimental values\cite{Perdew_DFT,Perdew1983,ShamDFT,PaulaDFT}. Predicting $E_g$ of semiconductors and insulators requires going beyond LDA or GGA approximations in DFT, making the calculations much more involved and computationally expensive.

\begin{figure}
\includegraphics[width=3.0 in]{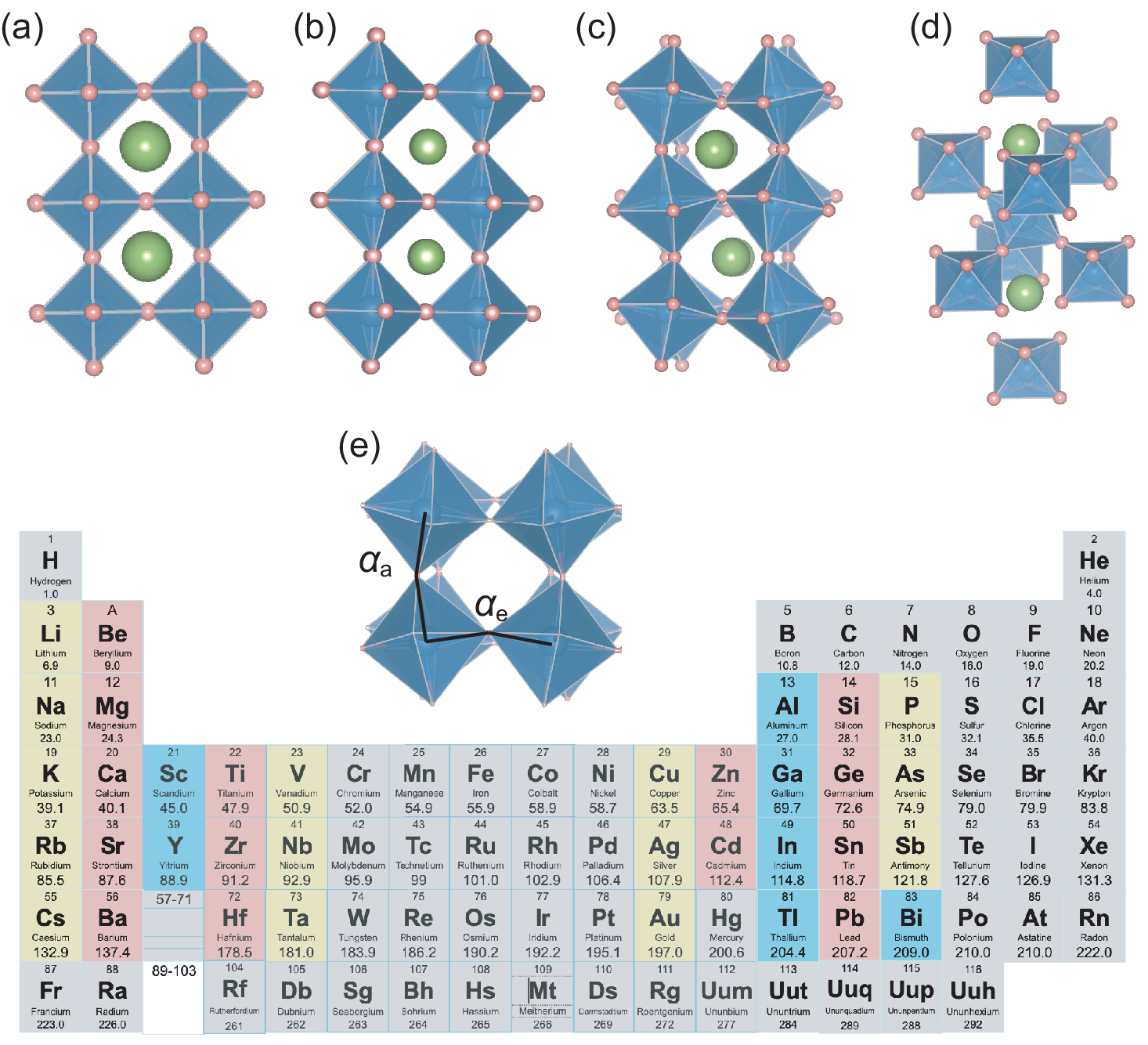}
\caption{Crystal structures of ABO$_3$ perovskite prototypes and selected A and B atoms. Crystal structure of (a) $Pm\bar{3}m$ cubic, (b) $I4_mmm$ tetragonal, (c)  $Pnma$ orthorhombic, and (d) $R\bar{3}c$ rhombohedral strutures of ABO$_3$ perovskites. Green, blue and red spheres represent A, B and O atoms, respectively. The apical and equatorial B-O-B bond angles, $\alpha_{a}$, and $\alpha_{e}$ are indicated in (e). The A and B atoms selected for this study are indicated in the Periodic Table in the lower panel.
}
\label{fig1}
\end{figure}

Methods that accurately predict bandgaps are very expensive with respect to both computational resources and wall time. The simplest approach is to mix Fock exchange with GGA exchange in a hybrid functional\cite{Becke1993,Perdew1996_1,Heyd2003,Heyd2006}, partially correcting the self-interaction error in DFT-LDA/GGA, giving bandgaps very close to the experimental values for many materials\cite{Brothers2008,Xiao2011,Kim2009,Henderson2011}. This increases the computation time tenfold compared to DFT-LDA/GGA calculations. More formally rigorous approaches would be to use the Green’s function quasi-particle GW \cite{Luoie1996,Kresse2006,Chen2015} or the wavefunction-based quantum Monte Carlo\cite{Hunt2018,Yang2020,Hunt2020} method, yet at the expense of at least an extra order of magnitude in computational time. As a result, these are not generally amenable to high-throughput computational approaches, posing a stringent obstacle to novel materials discovery. 

Machine learning (ML) techniques have emerged as powerful tools in materials science research, with applications in a variety of directions, such as prediction and classification of crystal structures\cite{Fischer2006,CARR2009339,Pilania2015,Yamashita2018,Ye2018,Rajan2018}
and building predictive models of various materials properties \cite{Medasani2016,Lee2016,Pilania2016,Bart2017}. 
Recent efforts also include predicting bandgaps, however with limited accuracy\cite{Zhuo2018,Lu2018,Xie2018,Allam2018,Olsthoorn2019}. A straightforward direction would be to predict bandgaps using the DFT-GGA band structures available in AFLOW database\cite{Curtarolo2012} as a training set for machine learning approaches. However, this would have limited use considering that the predicted bandgaps would still be severely underestimated. Or one could use DFT+U\cite{Anisimov1997} for bandgaps, with computational costs similar to those of DFT-LDA/GGA; the problem is what value of $U$ to choose and the justification of applying $U$ to dispersive valence and conduction bands. An interesting approach involves graph convolutional neural networks (CGCNN) based on atomic connections in the crystal structure after being trained using DFT bandgaps\cite{Xie2018}. However, this method was also trained and aimed at DFT-GGA bandgaps. Recently, reports on automated, high-throughput calculations of bandgaps based on hybrid functional have appeared in the literature\cite{Jie2019,Pilania2017,Ward2016,Huang2019}, pointing toward more reliable predictions of bandgaps, yet the nature and size of the band-gap corrections from the DFT-GGA values have not been discussed or analyzed.

\begin{figure}
\includegraphics[width=6in]{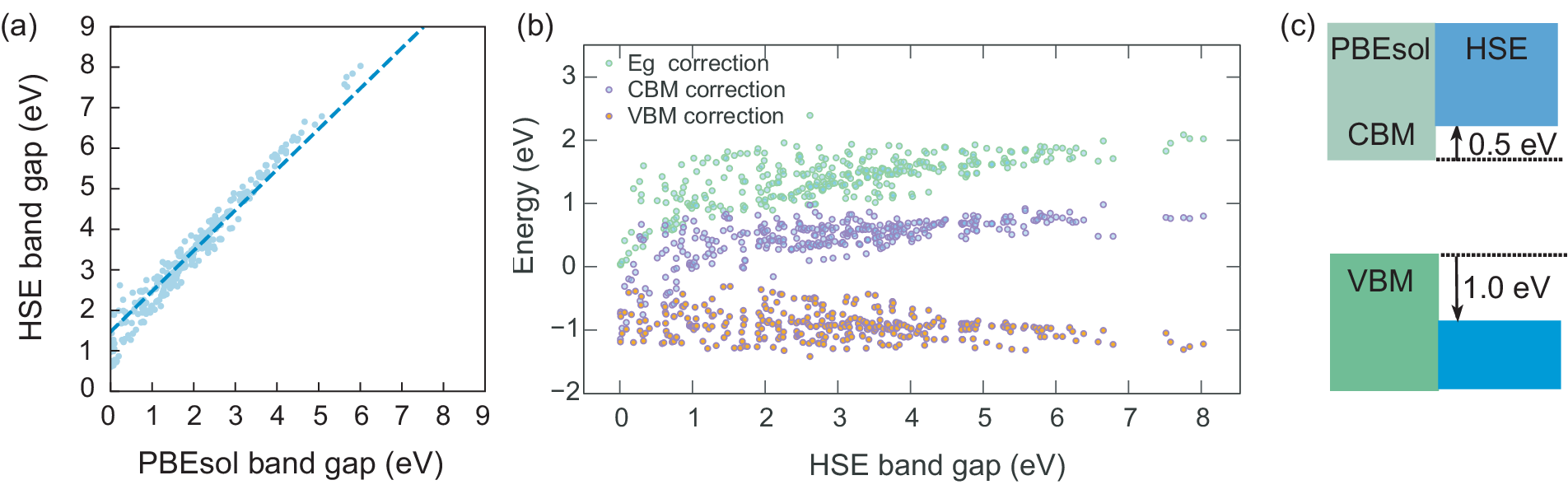}
\caption{Correction of the bandgap of ABO$_{3}$ perovskites based on HSE06 and DFT-GGA PBEsol calculations. (a) HSE06 vs PBEsol bandgaps, (b) the band-gap correction ($\Delta E_{g}$, light green), correction of the valence-band maximum $\Delta{\rm VBM}$ (light blue) and conduction-band minimum $\Delta{\rm CBM}$
(dark red) vs HSE06 bandgap. (c) schematic of the correction of the band-edge of the positions. The dashed line in (a), placed to guide the eye, has slope equal 1 and crosses the vertical axis at 1.5 eV. 
}
\label{fig2}
\end{figure}

In this work, we developed machine learning models for mapping bandgaps computed with DFT-GGA into bandgaps with higher accuracy HSE06 hybrid functional. We chose perovskite oxides as example to demonstrate the applicability of our approach. Oxide perovskites are a class of compounds that are of great importance in technology and basic sciences\cite{pervoskitea}, comprising semiconductors, insulators, ferromagnetic and antiferromagnetic, ferroelectric, multiferroic, piezoelectric, and high T$_{c}$ superconductor materials\cite{pervoskiteb}. The wide range of properties is often associated with the orbital character of the bands near the Fermi level and are strongly affected by variations in the crystal structure, such as octahedral rotations and distortions that are associated with deviations from the perfect cubic crystal structure\cite{pervoskitebc}. Accurate prediction of their electronic structure, bandgaps, and position of valence and conduction bands with respect to vacuum level, is crucial for designing novel devices. 
An interesting feature of ABO$_{3}$ perovskite semiconductors and insulators is the dependence of their band  gaps on the metal elements A and B as well as on rotations and tilting of the BO$_6$ octahedra. Here we restricted the scope of perovskite materials to those for which the valence band is derived from oxygen 2$p$ orbitals and the conduction band is derived from A or B valence orbitals, as indicated in Fig. 1. We did not consider perovskites where the valence and conduction bands are determined by transition metal $d$ orbitals and the gap associated with spin-splitting of $d$ bands or $d$-$d$ transitions. We explicitly included octahedral tilting and rotations leading to tetragonal, orthorhombic, and rhombohedral crystal structures as shown in Fig. 1. 
Using a high throughput approach\cite{Jain2015}, we calculated the band structures of the perovksites with PBEsol and HSE06 functionals. We analyzed the mapping of valence band maximum (VBM) and conduction band minimum (CBM) between PBEsol and HSE06 functionals by employing different machine learning models.
Our combined DFT-ML model predicts $E_{g}$ within an error of 0.16 eV to that of HSE computed  $E_{g}$, and reveals the main atomic and structural factors that determine the correction to the VBM, CBM, and consequently  $E_{g}$ predicted at GGA level.

\section{Methods}

The first-principles calculations are based on DFT within the generalized gradient approximation of Perdew, Burke, and Ernzerhof revised for solids (PBEsol)  \cite{Perdew} and the projector augmented wave  method \cite{Blochl1994,Kresse1999} as implemented in the Vienna Ab initio Simulation Package (\textsc{VASP}) \cite{Kresse1993a,Kresse1993b}. The wave functions are expanded in plane waves with cutoff energy of 650 eV. Structure optimizations are performed using 7$\times$7$\times$7, 7$\times$5$\times$7, 7$\times$5$\times$5, and 7$\times$7$\times$7 $\Gamma$-centered $k$-point grid for the integrations over the Brillouin zones of the cubic, tetragonal, orthorhombic, and rhombohedral primitive cells, respectively. The screened hybrid functional  HSE06  \cite{Heyd2003,Heyd2006} is employed to compute target bandgaps, using the structural parameters found using the PBEsol functional. In tests we found that PBEsol and HSE06 give lattice parameters that differ by less than 1\%, and in good agreement with experimental values. So we neglected the differences in the bandgap calculated using the PBEsol-optimized lattice parameters and those calculated using the HSE06-optimized lattice parameters. Test calculations indicate that these differences are less than 0.1 eV.

We used different ML algorithms to build our band-gap prediction model, including the linear ridge regressor, kernel ridge regressor, and gradient boosted decision tree from open-source software package Scikit-Learn Toolbox\cite{scikit-learn}. The input to the model is comprised of atomic and structural properties, including the B-O-B apical angle $\alpha_{a}$ and B-O-B equatorial angle $\alpha_{e}$.
The regression fit to the input gives the predicted bandgaps. The prediction performance of the learning models is evaluated by the mean absolute error. 
The feature importance of all the descriptors is obtained with GBDT to interpret the importance of various descriptors in the training model.
We conducted a hyper-parameter search for GBDT models through grid search. The search parameters include max\_tree\_depth (1, 2, ... 10), number\_of\_estimators (50, 100, 150, ..., 1000), and learning\_rate (0.01, 0.02, ..., 0.2). We used the default hyper-parameter values in scikit-learn package for training LRR and KRR. We used MinMax Scaling to normalize the data for LRR and KRR. We did not normalize the raw features for training GBDT since normalization is not necessary to GBDT due to the tree-based model nature. We partitioned the data such that one-third of the data is reserved for testing. For the remaining two-thirds of the data, three-fold cross-validation (two-ninths of total data as the test set and four-tenths of total data as the training set at any given time) was used hyper-parameter tuning. Our MAE results are based on the testing data set.

\section*{Results and Discussion}

\begin{table}
  \centering
  \caption{Mean absolute error (MAE) used to evaluate the performance of fixed correction (FC), linear regression (LR), linear ridge regressor (LRR), kernel ridge regressor (KRR), and the gradient boosted decision tree (GBDT) models in predicting the corrections of the valence-band maximum ($\Delta {\rm VBM}$), conduction-band minimum ($\Delta {\rm CBM}$) and bandgap ($\Delta E_{g}$) of oxide perovskites in DFT-GGA PBEsol compared to the HSE06 values. }
    \begin{tabular}{cccccl}
    \hline\hline
         &  {FC}  &  {LR} & LRR   & KRR   & \ \ \ \ GBDT \\
    \hline
    $\Delta {\rm VBM}$ &  {$0.17$} &  {$0.09$}  & $0.10 \pm 0.01$ & $0.09 \pm 0.01$ & $0.09 \pm 0.01$ \\
    $\Delta {\rm CBM}$ &  {$0.24$} &  {$0.17$} & $0.19 \pm 0.01$ & $0.15 \pm 0.01$ & $0.10 \pm 0.003$ \\
    $\Delta E_{g}$ &  {$0.32$} &  {$0.21$} & $0.23 \pm 0.02$ & $0.20 \pm 0.01$ & $0.16 \pm 0.01$ \\
    \hline
    \end{tabular}%
  \label{tab:comp}%
\end{table}%

We selected 118 oxide perovskites ABO$_{3}$, and for each we considered four crystal structures, with symmetries $Pm\bar{3}m$ (cubic), $I4/mmm$ (tetragonal), $Pnma$ (orthorhombic) and $P63/mmc$ (rhombohedral), as shown in Fig. 1, totaling 472 structures. The selected A and B atoms, also indicated in the Periodic Table in Fig. 1, are: A = Li, Na, K, Rb, Cs, Cu, Ag, Au, Be, Mg, Ca, Sr, Ba, Pb, Zn, Cd, Sn, Sc,Y, La, or Bi, and B = P, As, Sb, V,  Nb, Ta, Si, Ge, Sn, Ti, Zr, Hf, Al, Ga, In, or Tl, such that the considered compounds satisfy valence(A)$+$valence(B)=6. A data set of DFT-GGA bandgaps was constructed using this set of materials.

\begin{figure}
\includegraphics[width=6in]{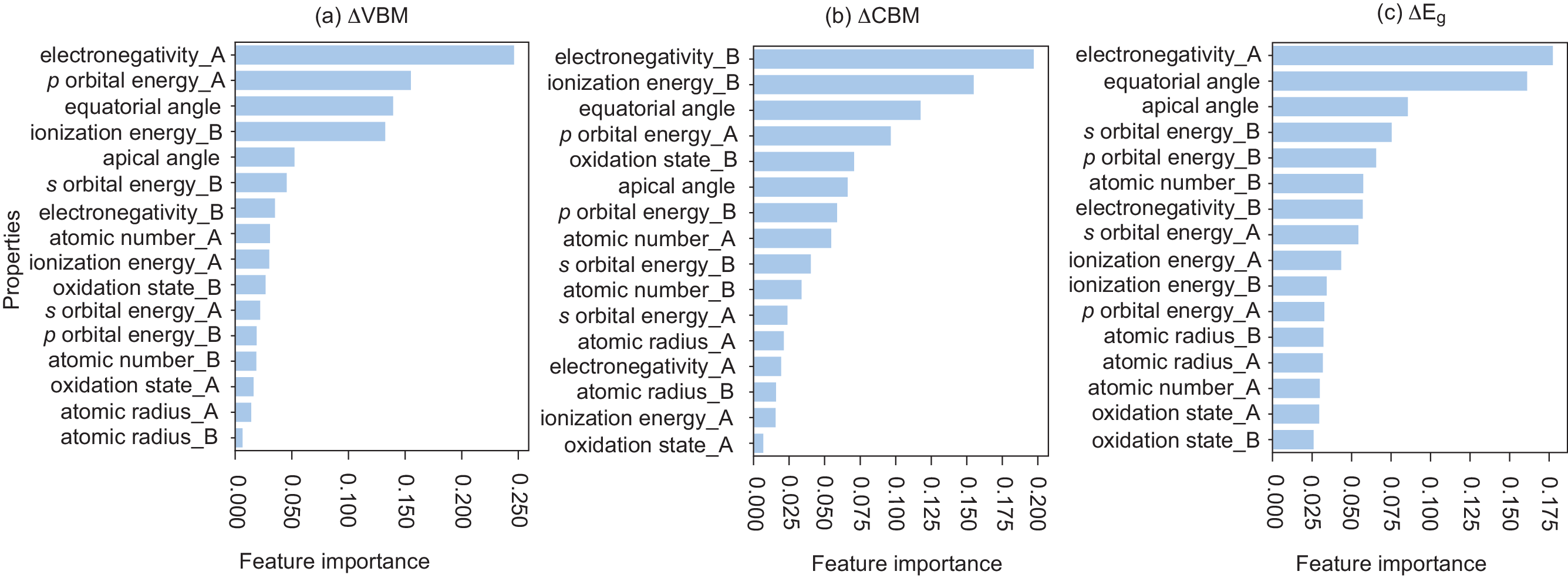}
\caption{Feature importance in the gradient boosted decision tree (GBDT) model for determining the band-gap ($\Delta E_{g}$) and band-edge corrections ($\Delta {\rm VBM}$, $\Delta {\rm CBM}$) of ABO$_3$ perovskites. }
\label{fig3}
\end{figure}

The four crystal structures for all ABO$_{3}$ compounds were first optimized with the DFT-GGA PBEsol functional. Then their electronic structures were calculated using PBEsol and HSE06. In this way, since the average electrostatic potential is used as the reference for the Kohn-Sham band energies, and does not depend on exchange and correlation, we can directly compare the PBEsol and HSE06 band structures, extracting the corrections for VBM and CBM, and the bandgap (i.e., $\Delta {\rm VBM}$, $\Delta {\rm CBM}$, and $\Delta E_{g}$).  We note that for all compounds studied here, the VBM for the cubic structure occurs at the R point (0.5, 0.5, 0.5) and the CBM occurs at the $\Gamma$ point in the cubic Brillouin zone, characterizing an indirect R-$\Gamma$ fundamental bandgap.  For the tetragonal, orthorhombic, and rhombohedral structures, both VBM and CBM occur at $\Gamma$, characterizing a direct $\Gamma$-$\Gamma$ fundamental bandgap. 

The calculated HSE06 bandgaps vs PBEsol bandgaps are shown in Fig. 2(a) 
and Supplementary Information. There are 383 data selected in the 472 materials since others are not stable according to the to DFT calculation. First, we note that the HSE06 predicted bandgaps have a nearly linear relationship with the DFT-GGA predicted bandgaps. We applied a simple linear regression fit using $y=ax+b$ between the two sets of bandgaps, and obtained $a=1.12$ and $b=1.15$. The resulting mean absolute error (MAE) is 0.21 eV, which is comparable to theMAE' s obtained with the more complicated models presented in the study. Since the value of $a$ is close to 1, the data had been fit to an even simpler model of fixed correction, $y=x+b'$. Fixed correction is very appealing due to its simplicity and provides an intuitive physical insight into the nature of the correction. The optimal $b'$ was found to be 1.5 eV with an MAE of 0.32 eV. The MAE of the fixed correction model compares well with the typical error in the DFT predicted bandgaps even when hybrid functionals are used. The fixed correction model implies that DFT-GGA underestimates the bandgap with respect to HSE06 by $\sim$1.5 eV. This is quite surprising given that in general DFT-LDA/GGA does not underestimate bandgap of semiconductors and insulators by a fixed amount\cite{Hinuma2014}. The largest deviation from this trend is observed for compounds containing Cu, Pb, and Sn occupying the A site. In the case of Cu-B-O$_3$ compounds, the Cu $d$ orbitals mix with the O 2$p$ orbitals, pushing the VBM to higher energies. In the case of Sn-B-O$_3$ and Pb-B-O$_3$, the VBM has large contributions from Sn and Pb $s$ valence orbitals, which also pushes the VBM to higher energies. In all the cases where the valence band is mostly derived from O 2$p$ orbitals, the approximate 1.5 eV band-gap correction fits the data quite well.  

The separated corrections $\Delta {\rm VBM}$ and $\Delta {\rm CBM}$, i.e., the amount the VBM and CBM in HSE06 differ from the VBM and CBM in DFT-GGA are shown in Fig. 2(b). Contrary to common wisdom, where it is often assumed that to correct the DFT-GGA bandgap only an upward shift of the CBM is necessary, we find that about 2/3 of the gap correction comes from shifting down the VBM and only about 1/3 of the correction comes from shifting the conduction band upward. This is attributed to large self-interaction correction of the O 2$p$-derived valence bands in these materials.  Again, the outliers, where the VBM is corrected by a lesser amount, correspond to compounds containing Cu, Sn, or Pb in the A site. It is also interesting to note the correction in the VBM derived from O 2$p$ is larger than the correction of CBM derived from $d$ orbitals, such as in SrTiO$_3$ and similar compounds, despite the rather flat nature of their conduction bands that are derived from the quite localized transition metal $d$ orbitals. Finally, we also note that the band-gap correction $\Delta E_{g}$ is slightly larger than 1.5 eV for compounds with larger bandgaps, approaching 2 eV, and this is traced back to the correction of the CBM which approaches 1 eV for compounds with $E_{g}$ is $\gtrsim$ 4 eV.

Having established the band-gap correction for these oxide perovskites, we now turn to machine learning techniques to develop a model that correlates the $\Delta {\rm VBM}$, $\Delta {\rm CBM}$, and $\Delta E_{g}$ corrections to atomic and structural properties of the compounds. The atomic properties as input to the machine learning models include electronegativity, ionization energy, valence-orbital energies, and atomic radius of both A and B atoms. Structural properties include octahedral tilting and rotations that are characterized by the apical $\alpha_{a}$ and equatorial $\alpha_{e}$ angles corresponding to B-O-B angles parallel and perpendicular to the $c$ axis. We employed three machine learning models, which are the linear ridge regressor (LRR), kernel ridge regressor (KRR), and the gradient boosted decision tree (GBDT) regressor, as implemented in Scikit-Learn Toolbox\cite{scikit-learn}. We used a regularization strength of $0.01$ to both LRR and KRR models. For the KKR method, we used a polynomial kernel with a maximum order of $3$. For the GBDT model, we set the maximum tree depth to $5$ with $500$ base estimators.

The prediction performance of the LRR, KRR, and GBDT models can be seen in 
Table.~\ref{tab:comp}. In these models, we use two third of the data as the training set. We also use mean absolute error (MAE) to measure the performance in predicting $\Delta {\rm VBM}$, $\Delta {\rm CBM}$, and $\Delta E_{g}$. Among the three models, GBDT gives the highest prediction accuracy with low variance; the KRR model performs better than LRR.  Note that we obtain lower MAE than previous models\cite{Zhuo2018,Gladkikh2020,Pilania2017,Rajan2018,Mishra2019}, likely to the better quality or more uniformity of our training dataset. The results indicate that there exists a nonlinear relation between the input properties and the target results, explaining why the pure linear model LRR performs poorly.  Note that all three ML models predict $\Delta {\rm VBM}$ with similar performance, indicating that the VBM correction has a more linear relationship with the input properties than the CBM and $E_g$ corrections.

What are the main atomic and structural properties that determine the band-gap and band-edge corrections? The answer is shown in Fig. 3, where the input properties are ranked according to their contributions to the prediction accuracy based on the mean decrease in the impurity of the GBDT model~\cite{breiman2001random}. We find that the electronegativity, the energy of $p$ valence orbital of atom A, and the equatorial angle of the octahedral rotation are the main properties that determine $\Delta {\rm VBM}$. For $\Delta {\rm CBM}$, the main properties are the electronegativity, ionization energy of atom B atom and the equatorial angle of the octahedral rotation. For more advanced feature importance evaluation methods with higher local and global consistency and interpretability, we refer to the literature\cite{lundberg2017unified, lundberg2018consistent}. We have also applied LRR, KRR and GBDT models to the data by excluding the discovered less-important features for each label, and no obvious accuracy improvement is identified.

For both $\Delta {\rm VBM}$ and $\Delta {\rm CBM}$, the equatorial angle determines the overlap between the orbitals of B and O in the directions parallel to the $a$-$b$ plane, which in turn, affect both VBM and CBM positions. Note that the dependence on the apical angle $\alpha_a$ is less than that on the equatorial angle $\alpha_e$ since the former affects the B-O orbital overlap only along the $c$ direction.  Finally, we also note that the relative importance of the electronegativity, ionization energies, and rotation angles is higher for atom A than for atom B in determining the bandgap.  This is attributed to the larger contribution of the VBM correction than the CBM correction to $\Delta E_{g}$.

\section*{Summary}

Using high-throughput DFT-GGA PBEsol and HSE06 calculations we determined the bandgap correction of a representative set of oxide perovskites, finding that the HSE06-based correction pushes down the valence band by $\sim$1 eV and pushes up the conduction band by $\sim$0.5 eV. These results are then used in machine learning models that include atomic and structural properties as input to determine the corrections to the valence band, conduction band, and bandgap. The properties used as fitting parameters are ranked according to their relative importance to the corrections. We find that electronegativity of the A and B atoms together with the equatorial angle of rotation of the BO$_6$ octahedra are the main factors involved in the corrections. These results serve as starting point and guide to developing machine-learning-based approaches applicable to the discovery of novel electronic materials.

\section*{Data availability}

The datasets generated and/or analysed during the current study are available in
the GitHub repository \url{https://github.com/vera-weili/perovskite_ML}

\section* {Acknowledgements}
This work was supported by the National Science Foundation EAGER-1843025. 
This research was also supported by the Xtreme Science and Engineering Discovery Environment (XSEDE) facility, National Science Foundation grant number ACI-1053575, and the Information Technologies (IT) resources at the University of Delaware, specifically the high-performance computing resources. This research was also supported in part by the following National Science Foundation grants: 1447711, 1743418, and 1843025. AJ acknowledges support from NSF Faculty Early Career Development Program DMR-1652994.
Wei Li was partly supported by the Laboratory Directed Research and Development Program of Los Alamos National Laboratory (LANL) under project number 20210087DR. LANL is operated by Triad National Security, LLC, for the National Nuclear Security Administration of U.S. Department of Energy (Contract No. 89233218CNA000001).

%
%

\end{document}